\documentclass[superscriptaddress, amsmath,amssymb,prx,twocolumn,]{revtex4-2}

\usepackage{graphicx}% Include figure files
\usepackage{dcolumn}% Align table columns on decimal point
\usepackage{bm}% bold math

\begin{document}

\title{The Quantum House Of Cards}

%\author[a,1]{Xavier Waintal}
%\affil[a]{Universit\'e Grenoble Alpes, PHELIQS, CEA, Grenoble INP, IRIG, Grenoble 38000, France}
%\leadauthor{Waintal}
%\authordeclaration{The author declares having no competing interests.}
%\correspondingauthor{\textsuperscript{1}To whom correspondence should be addressed. E-mail: xavier.waintal@cea.fr}
%\keywords{Quantum computing $|$ Quantum error correction $|$ Quantum algorithms $|$}

\author{Xavier Waintal}
\affiliation{Universit\'e Grenoble Alpes, PHELIQS, CEA, Grenoble INP, IRIG, Grenoble 38000, France}
\email{xavier.waintal@cea.fr}

%\dates{This manuscript was compiled on \today}
%\doi{\url{www.pnas.org/cgi/doi/10.1073/pnas.XXXXXXXXXX}}

\begin{abstract}
Quantum computers have been proposed to solve a number of important problems
such as discovering new drugs, new catalysts for fertilizer production, breaking encryption protocols, optimizing financial portfolios, or implementing new artificial intelligence applications. 
Yet, to date, a simple task such as multiplying $3$ by $5$ is beyond existing quantum hardware. 
This article examines the difficulties that would need to be solved for quantum computers to live up to their promises. 
I discuss the whole stack of technologies that has been envisioned to build a quantum computer from the top layers
(the actual algorithms and associated applications) down to the very bottom ones (the quantum hardware, its control electronics, cryogeny, etc.) while not forgetting the crucial intermediate layer of quantum error correction. 
\end{abstract}

\maketitle
%\thispagestyle{firststyle}
%\ifthenelse{\boolean{shortarticle}}{\ifthenelse{\boolean{singlecolumn}}{\abscontentformatted}{\abscontent}}{}

%%%%%%%%%%%%%%%%%%%%%%%%%%%%%%%%%%%%%%%%%%%%%%%%%%%%%%%%%%%
\subsection*{\label{sec:intro} Introduction}
%%%%%%%%%%%%%%%%%%%%%%%%%%%%%%%%%%%%%%%%%%%%%%%%%%%%%%%%%%%

I am very skeptical that a quantum computer will ever solve serious problems.
When I express these doubts to colleagues, the answer I mostly get is a variation along the line of the following: ``You're right, it looks difficult. But when the first transistor was built, one could never have foreseen the computer revolution, internet, smartphones, AI... Quantum bits might be the new transistor.'' This is a very compelling argument. In fact, some qubits {\it are} actually made of transistors \cite{piot2022}. Such analogies can be very striking, yet they can also be very deceptive as we are naturally inclined to use the one that fits our purpose best.
Other analogies compare quantum computing with the early days of aviation or the qubit with the brick with which we eventually built skyscrapers.
I find that these analogies miss a crucial difference between quantum computing and earlier technologies: the brick, the airplane, and the +1 Volt/ -1 Volt state of the transistor are all intrinsically robust. Qubits are much closer to a deck of cards
than to a brick.
 Could we have built our skyscrapers with cards? Probably not, the house of cards technology lacks robustness. 
 Quantum technologies are also, almost by definition, the opposite of robust; they rely on subtle, volatile, transient physical effects.  They intend to prevent a macroscopic object from behaving classically. Every single player in the game, be it a vibration, an electromagnetic mode, or a nearby charge wants the quantum computer to behave like a classical object; this is decoherence \cite{haroche1996}. Trying to build a quantum computer means picking up a fight with the second law of thermodynamics. The entropy of the system cannot be allowed to increase or the quantum state is simply gone.

\begin{figure}
\begin{center}
\includegraphics[width=0.95\linewidth]{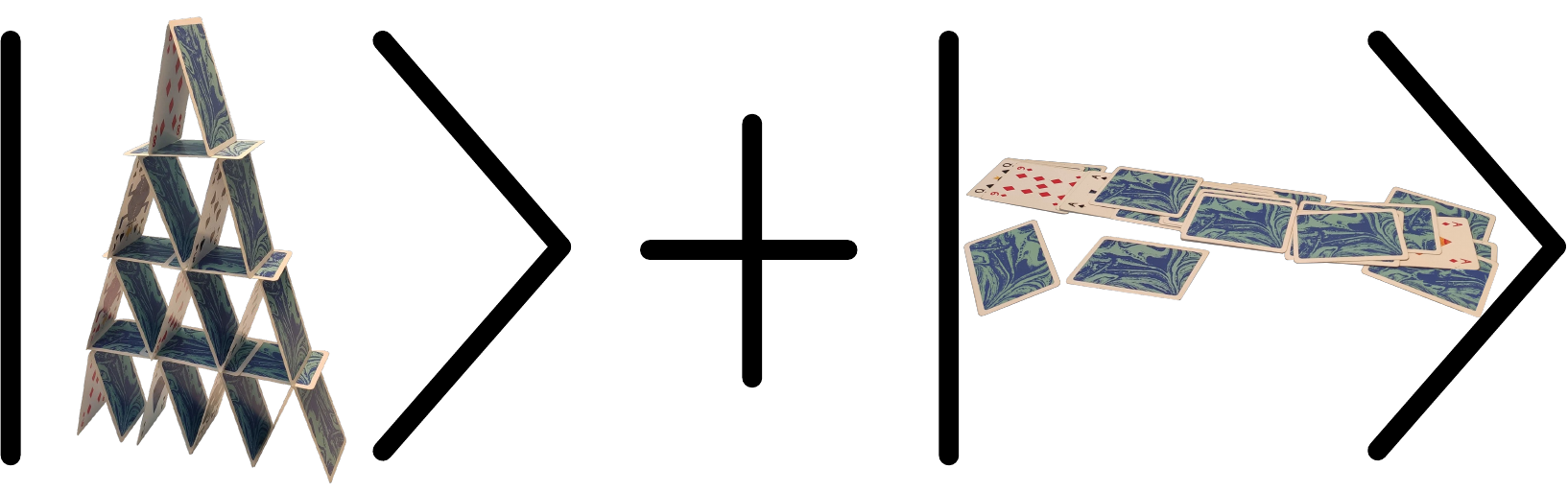}
\caption{A quantum computer internal state is a macroscopic quantum state
described by an exponentially large set of complex numbers. Such states are subject to decoherence and very fragile.
}
\label{fig:QhC}
\end{center}
\end{figure}

The goal of this article is to dispel a few myths around quantum computers, and coldly examine what it would take to build one. I will point out the different challenges that lie on the path to building a quantum computer, the leading one being the exponential decay of the fidelity due to decoherence. I will also discuss the envisioned applications of quantum computing and the extraordinary constraints that they put on  quantum hardware. As far as I am aware, the statements written below are not particularly controversial, at least to the scientific experts in the field. Depending on the reader's background, some of these statements might be obvious, some of them new. Taken together, they provide a global picture that should help one to make an informed opinion on the main question of the field: Will it work?

As advertized in the first sentence, the author of this review is very skeptical about the prospects of quantum computing. It should be stressed, however, that this skepticism does not extend to quantum physics in general or to other aspects of quantum technologies. The revolution going on in experimental quantum physics is genuine. What is unclear, in my opinion, is the possibility to use this progress to perform new types of computation.  

%%%%%%%%%%%%%%%%%%%%%%%%%%%%%%%%%%%%%%%%%%%%%%%%%%%%%%%%%%%
\subsection*{\label{sec:fidelity} The fundamental law of analog machines}
%%%%%%%%%%%%%%%%%%%%%%%%%%%%%%%%%%%%%%%%%%%%%%%%%%%%%%%%%%%

At the root of this analysis is the simple observation that a quantum computer is an analog machine \cite{dyakonov2014}, i.e. its internal state is described by a large set of complex numbers that can vary {\it continuously}. This is in contrast to digital machines, like our desktop computers, whose internal states are described by bits that can take two (macroscopically distinct) values: zero and one. Microelectronics experts know very well that it is very difficult to perform long calculations with analog machines because every operation is executed with a {\it finite precision}. The same limitation applies to operations on quantum bits: For instance, we might want to rotate one by 40 degrees around the z-axis, but instead end up doing a $40.1$ degree rotation around a slightly tilted axis. These errors accumulate, and eventually the internal state of the quantum computer becomes blurred. 

We measure the global precision that we have achieved with a quantity called the fidelity $F$ \cite{ayral2023}. One can think of the fidelity as the probability for the result of the full calculation to be correct. The fidelity decreases exponentially as,
\begin{equation}
\label{eq:fidelity}
F \approx e^{- \epsilon_0 N_0 - \epsilon_1 N_1- \epsilon_2 N_2 - \dots }  
\end{equation}
where $\epsilon_1$ is the error per one qubit operation and $N_1$ the number of such operations. Similarly,  $\epsilon_2$ and $N_2$ refer to the error level for operations that couple two qubits. There exist also other error channels (e.g. measurement errors, leakage errors, etc.) not discussed here. Interestingly, the phenomena of decoherence occur even when we do nothing, so that we have to introduce the error level $\epsilon_0$ for the $N_0$ idle times as well. Physically, $\epsilon_0 N_0$ is essentially given by $\epsilon_0 N_0 \approx n T/T_2$ where $n$ is the number of qubits, $T$  the duration of the calculation and $T_2$ the phase coherence time of the quantum hardware. Since $T$ scales at least as $n$ for non-trivial calculations (much faster in the case of \emph{e.g.} Grover algorithm), it follows that the fidelity decreases at least as fast as $F \sim \exp(- a n^2)$ ($a$ is a constant). The error levels $\epsilon_1$ and $\epsilon_2$ include, in addition to decoherence, the errors due to the act of manipulating the qubits such as the finite precision of the amplitude and duration of the microwave pulses or the effect of \emph{e.g.} crosstalks. There is a tremendous level of evidence, both theoretical and experimental, for the validity of \eqref{eq:fidelity}; it is actually used to benchmark the quality of the qubits. 
Perhaps the most compelling validation so far of \eqref{eq:fidelity}
is the seminal Google experiment of ``quantum supremacy'' \cite{arute2019}.
Indeed, if one sets aside the supremacy claim, what this experiment really does is establishing that \eqref{eq:fidelity} actually holds for a rather large system of $n = 53$ quantum bits.
Typical values shown by the best existing hardware are around 
$\epsilon_1\approx 0.0001$ and $\epsilon_2\approx 0.002$.
The numbers $N_0$, $N_1$, $N_2$ depend on the type of application, and on the size of the problem, i.e. the number of quantum bits $n$ that will have to be used.
For a given problem, one may use \eqref{eq:fidelity} to evaluate the probability of success on given hardware.

%%%%%%%%%%%%%%%%%%%%%%%%%%%%%%%%%%%%%%%%%%%%%%%%%%%%%%%%%%%
\subsection*{The fruits are few and not hanging low}
%%%%%%%%%%%%%%%%%%%%%%%%%%%%%%%%%%%%%%%%%%%%%%%%%%%%%%%%%%%

There is a small paradox attached to quantum computing with respect to applications.
The way it has been presented to the public, there exists a large number of them, with dazzling implications. It has been claimed that quantum computers would help with food production by designing new catalyzers for fertilizer production; they would allow one to design new drugs or chemicals; they would play a major role in (quantum) machine learning; they would address cryptographic security problems\dots On the other hand, at the moment, only a handful of separate quantum algorithms exist. In fact, all the above mentionned applications use variations on these few algorithms. 
This is in sharp contrast to classical computing where it is relatively easy to design a new algorithm and there exists an almost unlimited number of them. The contrast remains even after taking into account the relative size and age of the two fields: there is no simple path to write a new quantum algorithm. Also, as we shall see, all known algorithms are very demanding in terms of hardware quality and quantity. 

The most famous one is Shor's algorithm that solves the factoring problem \cite{shor1994}. Given a large integer $c= a b$ of $n$ bits, it calculates the two factors $a$ and $b$. Its importance stems from the fact that, because this problem was believed to be exponentially hard, it is at the root of the RSA encryption system that is used throughout the internet. So someone in possession of a quantum computer could eavesdrop on internet communications that are supposed to be secure. Because this threat has been taken seriously,
new encryption protocols, proven to be quantum resilient, will soon be deployed to replace RSA. Considering only two-qubit gate error, $N_2 \approx 10 n^3$ for Shor's algorithm. Breaking RSA for realistic key sizes would imply factoring numbers represented by $n = 2048$ bits. From $F \approx e^{-\epsilon_2 10 n^3}$, we obtain that  one would need $\epsilon_2 \le 10^{-11}$ for the algorithm to succeed with a reasonably high probability. This would require an improvement by a factor of one billion with respect to the current experimental state of the art. In the particular case of Shor's algorithm, the effect of noise can be put onto firm mathematical grounds \cite{cai2023}.

Second after Shor comes Grover. Grover's algorithm \cite{grover1997} provides a more modest speed-up than Shor's, at most quadratic from $2^n$ down to $\sqrt{2^n}$ operations. 
It is however widely popular because in contrast to Shor's algorithm, which is highly specialized, Grover can be applied to a wide variety of problems including the difficult ``NP complete'' class. 
Its applicability is however questionable \cite{stoudenmire2023} since it requires an exponentially large number of operations to complete, hence the fidelity decays as the exponential of an exponential: $F \approx e^{-\epsilon_2 n \sqrt{2^n}}$. 
Solving a $n=100$ instance (where it is expected to become faster than its  classical analogue in the most favorable situation) would require $\epsilon_2 \le 10^{-17}$ not to mention an uninterrupted computation of thousands of years \cite{stoudenmire2023}. 
Using existing technologies, it is actually not possible to apply Grover reliably on just $5$ qubits. 
A recent attempt that used advanced noise mitigation techniques (not applicable to large systems) obtained the correct answer with just $15\%$
probability \cite{pokharel2022}. Beyond the difficulty to run Grover's algorithm in practice, a widely overlooked aspect is that a quantum advantage can only be obtained for the most {\it unstructured} problems. Real life problems almost always possess a lot of structure that classical algorithms take advantage of for dramatic speed-up \cite{stoudenmire2023,ayral2023}. Many proposals for quantum advantage are rather naive in that respect.

The third class of algorithm does not provide a quantum speed-up in theory,
but the hope is that it might provide one in practice. 
Among the possible applications, one that has attracted a very large interest is the possibility to construct variational ansatz to solve problems arising in quantum chemistry, e.g. calculate the activation energy for a chemical reaction and the influence of potential catalysts. 
The gate count $N_2$ is directly linked to the expressivity of the ansatz; in order to reach a sufficient (so called ``chemical'') accuracy, one needs to include at least excitations with three electron-holes \cite{eriksen2020} (which translates into $N_2\sim n^6$) and a large fidelity $F \ge 0.999$. For a $n=30$ electron problem accessible to most classical techniques, one arrives at $\epsilon_2 \le 10^{-12}$ \cite{louvet2023}.

The above examples are actually very generic: to be useful, a quantum computer must be able to employ rather deep quantum circuits. This statement can actually be put on firm grounds since shallow circuits can be simulated on classical computers very efficiently \cite{zhou2020,ayral2023}. In return, this means that to be useful
a quantum computer must present very low error levels to keep the success probability of the computation close to one. To most physicists the numbers above speak by themselves; it is not realistic to imagine this level of precision to be reached.  Suppose one is playing darts with a 1m wide target. Putting several darts within a circle of 5 cm of diameter is accessible with a little work. Putting the darts inside the 2 mm center requires a lot of impressive work 
(this is where we are now, $\epsilon_2= 0.002$). For the above quantum algorithms to work, one needs to put all the darts within a $10^{-11}$ m wide target.

All attempts to use quantum hardware for
applications have been so far restricted to very small problem sizes in order to evade the precision issue. 
For instance, the quantum circuit to perform $3\times 5$ is still beyond the reach of existing chips and so is the quantum circuit to factorize $15$. Indeed, in the seminal article where such a factorization has been advertized, the classical logic was pre-calculated on a classical computer in order to maintain the gate count at a low enough level \cite{vandersypen2001}. Also, the above estimates assume perfect connectivity between all qubits. 
In practice, this is never achieved except for small systems and one should therefore expect additional overheads due to the need of shuffling qubits around.

Finding applications where quantum computers could have a genuine advantage is difficult. However, there is a simple criterion that allows one to state what they {\it won't be able to do} \cite{hoefler2023}. Under very optimistic resource usage assumptions, it takes a quantum computer hours or days to arrive at a solution that would be challenging for a classical computer \cite{reiher2017,beverland2022}. Since this solution encompasses only a few tens of bits, it means that the information throughput that such a quantum computer would deliver is a few tens of bits per hour which translates into less than a milli-byte per second, i.e. twelve orders of magnitude less than what any laptop routinely provides (gigabytes per seconds). In other words, quantum computers deliver very few bits of information. These might be very useful bits but still they will be very few, so quantum computers will always be restricted to highly specific applications; they will not replace classical computers. For instance a quantum computer could in principle be used to calculate the binding energy of a molecule (this can be stored in few bits) but will not be able to generate an image (this would require millions of logical qubits). 
Conversely, applications that require large inputs to be given to quantum computers, for instance an image for a quantum machine learning application, will require at least as large a gate count to load this information into the quantum hardware hence automatically require ultra low noise levels to keep the fidelity high enough.

%%%%%%%%%%%%%%%%%%%%%%%%%%%%%%%%%%%%%%%%%%%%%%%%%%%%%%%%%%%
\subsection*{The duality of good coupling vs. low decoherence}
%%%%%%%%%%%%%%%%%%%%%%%%%%%%%%%%%%%%%%%%%%%%%%%%%%%%%%%%%%%

On the physics side, every field of quantum physics has proposed one or more platforms
on which one could try to realize a quantum computer. These different approaches can roughly be put on a scale from the ones best coupled to their environment (and hence can be manipulated efficiently but also suffer more from decoherence) to the least coupled ones (showcasing excellent coherence times but slow to manipulate). It is difficult to make an exhaustive list of what is being tried but one should keep in mind that optimizing one aspect is often done at the expense of another one and that the advantages of one approach often end up in being also a disadvantage. 
For instance, semiconducting approaches based \emph{e.g.} on the spin of a single electron can provide very small qubits, hence the possibility to scale up to large systems. 
They also benefit from the existence of proven fabrication technology.
Yet their smallness also means that they are more sensitive to defects at the atomic scale, and therefore display high variability. For instance, a good CMOS technology has around $10^{11} {\rm cm}^{-1}$ charges trapped at the silicon-oxyde interface of a transistor. This translates into variations of electric potential of several meV, not much smaller than the potential induced by the controlling electrostatic gates. Transistors that would be considered as nominally identical for microelectronics end up showing very large variations when used for qubits \cite{martinez2022}. So far, only very limited quantum states have been demonstrated with this approach.
Atomic physics based trapped ions have very small variability (all atoms are identical, only their environment may fluctuate) and long coherence. 
Yet they are several orders of magnitude slower than their semi-conducting counterparts which almost disqualifies them for a full scale implementation of quantum error correction (see below). Also, it is not clear yet how to scale up these systems succesfully. Photons have extremely long coherence times. Yet, as they do not interact directly, the two-qubit gates are obtained by measuring auxiliary photons (absorbed in a photodetector) and are intrinsically {\it probabilistic}. This means that one has, say, a $1/2$ probability of applying the gate and $1/2$ probability of doing something else. The probability to apply the intended gate sequence decreases exponentially as one increases the gate count, which is very bad.
Implementing an almost {\it deterministic} noisy two-qubit gate between photons requires using a large number of auxiliary photons and a highly complex set of interferometers and photodetectors just to catch up with what other platforms do naturally. Besides, while important progress has been made in solid state photon sources, their yield and indistinguishability are at the $\epsilon\sim 0.1$ level, still far from what would be needed. Other platforms, such as Rydberg atoms or electronic flying qubits \cite{bauerle2018} are
emerging at various positions on the ``coupling to the environment'' spectrum. As of now, superconducting circuits seem to be a good compromise with a state of the art at around a hundred qubits and $\epsilon_2 \approx 0.001$. Notice that with this level of error, only a fraction of the qubits can be used reliably if one wants to retain a reasonably high fidelity.

Each field of quantum physics offers a rich variety of different sub-possibilities.
For instance, within superconducting circuits \cite{kjaergaad2020}, early designs included the quantronium after which came the transmon (now the de facto standard) but other designs include the fluxonium (with state of the art characteristics), the gatemon or the bosonic qubits (that aim at implementing, at least partially, quantum error correction at the hardware level).

Extrapolating the pace at which progress is made is not straightforward.
Progress is often linked to the advent of disruptive ideas (e.g. the Cooper pair box, using a resonator as a harmonic oscillator, the transmon); the optimization part has been greatly accelerated by the large increase of resources devoted to quantum computing in the last few years with some research teams growing from a few persons to almost a thousand. The first demonstration of a ``qubit'' (i.e. coherent Rabi oscillations) in a quantum circuit was made in 1999. The first two-qubit gate was made in 2009 with $\epsilon_2\approx 0.1$ which was later improved to $\epsilon_2\approx 0.01$ and reached $\epsilon_2\approx 0.001$ in 2020  (see Table I of \cite{kjaergaad2020} for references and details). At the time of this writing, 
IBM has announced that its chip can sustain the $\epsilon_2\approx 0.001$ level with more than a hundred qubits \cite{kim2023}.

%%%%%%%%%%%%%%%%%%%%%%%%%%%%%%%%%%%%%%%%%%%%%%%%%%%%%%%%%%%
\subsection*{The ``salvation is beyond the threshold'' myth}
%%%%%%%%%%%%%%%%%%%%%%%%%%%%%%%%%%%%%%%%%%%%%%%%%%%%%%%%%%%

The contrast between the fidelities displayed by the quantum hardware and the requirements of the algorithms raises concerns about feasibility. The theoretical solution to this problem is an elegant mathematical construction, the quantum error correction (QEC). In QEC, one performs
two things: first, one encodes a {\it logical} qubit into $n_c>1$ physical ones so that no simple physical process would directly couple the logical zero with the logical one. Second, one repeatedly performs {\it partial} quantum measurements that leave the logical qubit unchanged but force the system to stay in the intended part of the (now enlarged) Hilbert space \cite{nielsen2000}. A widely spread idea, backed up by some strong mathematical theorems, is that once the error levels fall below a certain threshold $\epsilon_{\rm th}$ (that happens to be of the same order of magnitude as the current error level, $\epsilon_{\rm th}\approx 0.01$ for the best QEC codes), it becomes sufficient to increase $n_c$ to obtain an exponential decrease of the error level $\epsilon_L$ of the logical qubit. QEC is often summarized with statements of the form ``We can anticipate that analog quantum simulators will eventually become obsolete. Because they are hard to control, they will be surpassed some day by digital quantum simulators, which can be firmly controlled using quantum error correction.''\cite{preskill2018} or ``so, once the desired threshold is attainable, decoherence will not be an obstacle to scalable quantum computation.''\cite{divincenzo2000}
In short, QEC would turn an analog machine into an essentially digital one. However, QEC is itself entailed with considerable difficulties that I now detail.

The first difficulty is that QEC is associated with important overheads in terms of 
number of qubits and number of operations. I refer to this as the Russian doll nesting problem. This name reflects the fact that manipulating logical qubits involves several levels of nestedness: operations within operations within operations. Suppose that one wants to perform a simple rotation of one logical qubit by an angle $2\pi/64$ around the $z$-axis (a typical rotation occuring e.g. in the quantum Fourier transform, a key component of Shor's algorithm). To perform this rotation on 
a physical qubit is deceptively easy: one simply needs to wait for the time it takes for
the qubit to do this precession naturally. However, no such simple path can exist within QEC since, by construction, the logical zero and one are widely different states and since we want this rotation to be performed with arbitrary precision. Performing this simple rotation in what some have called a ``digital way'' (that is with an arbitrary high control on its precision), one decomposes it into several nested levels of operations. At the first level of the Russian doll, the rotation is decomposed into a product of a few ultra-precise elementary operations. This is similar to approximating a real number with a fraction of two integers, the more precise one wants to be, the larger the integers or, in our case, the larger the number of elementary operations. Some of these elementary operations will be relatively straightforward to implement, for instance they could amount to applying the same operation to each of the $n_c$ qubits. At least one of them will be hard however. In the most studied
QEC code (also one of the most robust), the so-called surface code \cite{fowler2012}, the hard gate is the $\pi/4$ rotation along z, called $T$. In a second level, one needs to fabricate special $\pi/4$ states very precisely. This is done through a process called ``magic state distillation'' that is planned to occur in ``$T$ gate factories''. Magic state distillation is itself a sort of QEC code that possesses a fixed point that is precisely the intended state. Since it is done with logical states it is therefore a sort of QEC of QEC code. One or more iterations of the distillation are necessary before the required accuracy is achieved. To perform the distillation, one must entangle different logical qubits with each other. This is achieved within the third nested level of the Russian doll using a process
known as ``braiding'' where the first logical qubit makes a complete loop around the second logical qubit. This loop is itself discretized in the fourth level of the doll where each discrete move of the qubit is intertwined with ``syndrome measurements'': a small quantum circuit ran on the entire chip at every clock cycle to perform the partial measurements mentioned above. Needless to say, the Russian doll nesting problem transforms the initially simple process of a one qubit gate into thousands, tens of thousands or more, physical operations depending on the required level of precision. It is interesting to note that $T$ seems to be a necessary evil. Indeed, it can be shown that quantum circuits that do not have the $T$ gates can be simulated easily, in polynomial time on a classical computer \cite{gottesman1998}.
The Russian doll bottleneck excludes all but the fastest hardware for practical applications. Even assuming gates as fast as $10 ns$, the overall computational time found for a full application often reaches days for a single answer to a single problem.
It also imposes stringent constraints on the scalability as millions of qubits (billions once one includes the $T$ gates factories) will be required \cite{reiher2017}.

The second difficulty I call the ``syndrome bottleneck''. At every single clock cycle (say every $\mu s$ for state of the art superconducting hardware), quantum error correction requires partial measurements of the physical qubits in order to stabilize the logical qubits. The results of these measurements are the ``syndromes'' that need to be tracked and analyzed in real time in order to reconstruct what type of error may have occured and modify the rest of the circuit accordingly. There are (almost) as many syndromes as there are qubits. A single error in the analysis of the syndrome jeopardizes the entire calculation. And the analysis of the syndrome is a difficult (NP-complete) problem.
For a billion physical qubits quantum computer, the data rate coming from the syndromes is
of the order of $10^{15}$ bits per second ($=10^9/10^{-6}s$). Even assuming that very few operations per bit are needed for the analysis, this means that a petaflop of computing power is required
solely for the purpose of syndrome analysis. This is the computing power of a full scale supercomputer. Even more worringly, there is a need to extract the $10^{15}$ bits per second  from the quantum chip, i.e. the equivalent of what goes through one million gigabit ethernet cables.

The last difficulty arises from the fact that not all errors have the same status. Indeed, a QEC code corrects {\it certain} types of errors. Others are not corrected and are fatal to the calculations. In the context of the surface code, the logical error takes the form
\begin{equation}
\epsilon_L \sim \left(\frac{\epsilon_2}{\epsilon_{\rm th}}\right)^{\sqrt{n_c}}
+ \epsilon_{\rm nc} n_c
\end{equation}
where $\epsilon_{\rm nc}$ is the error level for non-correctable errors \cite{waintal2019}. Hence increasing $n_c$ will only improve the logical error until the non-correctable errors start to dominate. For each set of errors, one can probably build a code that corrects them all (but the more errors, the more complex the code and the worse the threshold). However, for a given code there will always be non-correctable errors. In QEC terminology, such errors are referred as correlated errors. However they are correlated only from a QEC point of view. From a physics standpoint they are just regular physical processes (for instance, a global fluctuation of magnetic field due to fluctuations of the current in the coil qualifies). Some of these errors have already been identified. Examples include the leakage of a qubit outside of the intended computational space, a nearby two-level system that creates a memory effect \cite{spiecker2022} or, more dramatically, a collision with a cosmic ray that appears to impair superconducting chips \cite{mcewen2022}. Fortunately in the latter case, it seems that burying the quantum computer under 1.4km of non-radioactive stone alleviates the problem \cite{gusenkova2022}. 
The bottom line is that if and when one starts to use QEC, one will only be able to obtain {\it some} gain in precision, not an {\it arbitrary} gain. After that, one will be dominated by a non-correctable error, hidden behind a large flow of more mundane ones. One will need to find a way to isolate and then address this error (most probably at a cost for the threshold) and only then one will be able to increase $n_c$ again, until one will be hit by the next type of error. Given the precision targeted in the applications ($\epsilon_L \le 10^{-15}$ once one takes into account the QEC overhead), it is not clear how far one will be able to go in practice.

Recent attempts \cite{krinner2022,acharya2023} to construct a surface code with superconducting qubits have shown a mitigated success. Up to $n_c=25$ has been considered. However,
so far only the fourth level of the Russian doll nesting problem has been implemented, so that the logical qubit could not be manipulated. Also the logical qubit is still not as
good as the original physical qubits from which it is constructed. A real-time treatment of the syndromes has not been attempted so that syndrome analysis is done post mortem. This is fine in a proof of principle experiment when one can post-select some experiments, but not usable for a real quantum computation. Last, in one of these experiments \cite{krinner2022}, significant leakage occurs so that non-correctable errors can already not be ignored. The most complete attempt to implement all the stages of the Russian doll probably belongs to the trapped ion platform \cite{postler2022} using one of the simplest QEC codes (the $n_c=7$ Steane code). One logical $T$ gate was applied on one logical qubit. However the magic state distillation 
stage was bypassed, the code was still vulnerable to a single qubit error on the ancilla qubit, only a (partial) single cycle of syndroms was measured,  and it was analyzed post mortem through postselection. The result was a logical error $\epsilon_L > 0.1$, i.e. almost hundred time worse than the original single qubit error of the physical qubit. The best QEC gain so far was demonstrated for a single bosonic qubit in a 3D cavity where a gain of up to a factor two in the qubit error level was claimed with a logical $\epsilon_0 \approx 0.002$ per QEC cycle \cite{sivak2023,ni2023}.

\subsection*{The scaling up engineering challenge}

Constructing a genuine quantum computer, as it has been envisioned, is often described using the euphemism ``challenging''. The description above was meant to convey a more precise idea of the difficulties involved. To the fundamental problems outlined above, I will now add a few more mundane, almost trivial, ones.

Scaling up quantum technologies is an engineering problem that has its own set of exponential decays and other difficulties. 
Once again, optimizing one parameter is done at the expense of others, often the number of sustainable qubits in the architecture. For instance frequency crowding (several qubits whose frequencies lie within the bandwidth of each other) leads to correlated crosstalks (acting on one qubit triggers unwanted dynamics on other qubits). The optimization that work for a handful of qubits will not work any more once the number of qubits has been scaled up.  For instance, after increasing $n$, one won't be able to spread out their frequencies anymore and cross-talk will become more problematic.
 Another (exponential) problem is the yield of the fabrication. If a qubit has a probability $p$ to function nominally,
then the probability that all $n_{\rm tot}= n \ n_c$ qubits in the chip work is $p^{n_{\rm tot}}$. The value of $p$ must be {\it very} close to unity for this yield to be non-vanishing. Currently, it is rather common that several samples need to be fabricated in order for one to work correctly, so we see that this simple law imposes a considerable constraint on the reproductibility of the fabrication. 
This will be particularly important for solid state devices. 
Another crucial related problem is variability: some nominally identical samples may have slightly different microscopic environments that affect the qubit behavior or its coupling with others (such as an ill-positioned two-level system).
This is particularly problematic if this variability is not static: any drift in the system will lead to a loss in precision. In some technologies, the mere size of the chip might become an issue. For instance superconducting transmons are based on resonators that currently take up a surface of the order of 1 $mm^2$. 
Scaling up to a billion of them would require cooling down the surface of a one bedroom appartment to $10 mK$, which will be challenging. 
Besides the surface problem, existing dilution fridges offer typically $500 \mu W$ of cooling power at $100 mK$ ($\sim 1 W$ is available at $4 K$) for an electricity consumption of $\sim 10 kW$. This is sufficient for existing chips but would not survive the implementation of QEC. Another pressing problem is the control and addressing of all the qubits and the associated potential wiring/fan out bottleneck. The fan out problem is actually deeper  than it seems. Indeed, the number of control lines directly controls the amount of calibration that is made possible to mitigate a parasitic coupling or simply precisely calibrate a gate. One needs as many knobs as there are parameters to tune. Any attempt to compromise there will result in poorer noise level.

\subsection*{Conclusion}

I have tried to convey the idea that, perhaps, quantum computing as it has been envisioned so far is simply too difficult to happen. However, it remains that there is a genuine revolution that is going on in quantum sciences. We are exploring frontiers that were thought impossible only a few decades ago. Perhaps this is one of those cases where the journey is more important than the intended destination and that it is simply too early to foresee the applications.
After all, if we could not foresee the internet when we invented the transistor, perhaps we're not better off now?

There are some possible applications that I purposely did not include in this discussion. In particular, I did not address the so-called quantum simulations, adiabatic transitions or other analog systems.
Indeed, an important parameter of a quantum technology is its level of ``universality''. On one hand are systems which are fully universal; these are the systems that I described above and that would truly deserve the name of computers. At another extreme are systems that essentially ``simulate themselves'', albeit with some level of control onto the dynamics. Most of the difficulties outlined for universal (gate based) quantum computers do not apply there, or at least not as stringently. As we gain control of these systems and manage to implement dynamics that are of natural interest, they may provide us with important information about other less controllable systems such as new materials or important models such as the Hubbard model. This is perhaps a niche application but it might be a very important and fruitful one.

Also, it strikes me that some of the key applications that have been put forward for quantum computing (e.g. for solving chemistry problems or correlated matter) are slowly coming into the scope of classical methods. The simulation of the seminal quantum supremacy experiment, initially assumed by Google to take 10,000 years on the largest classical supercomputer \cite{arute2019}, now requires just 6 seconds according to their own new reckoning \cite{morvan2023} (an improvement of almost eleven orders of magnitude on the classical algorithm side). A similar fate befell the recent IBM quantum advantage experiment \cite{kim2023} which was simulated mere days after the article publication \cite{tindall2023,begusic2023,kechedzhi2023}.   
Indeed, a common misconception is that the many-body dynamics of a quantum computer is {\it necessarily} exponentially difficult to calculate on classical hardware. In reality it is {\it at most}
exponentially hard.  There are almost always hidden structures that scientists learn to exploit \cite{ayral2023} to speed up the calculations. The number of problems that were apparently exponentially difficult and that were solved in polynomial time with numerical techniques is quickly growing. A seminal example is the Kondo problem, solved with the Numerical Renormalization Group (NRG) for which Wilson was attributed the Nobel prize. In the last few years, cornerstone models such as the Hubbard model have started to yield to the convergent attack of Density Matrix Renormalization Group (DMRG) \cite{xu2023}, Quantum Monte-Carlo (QMC) \cite{xu2023}, Diagrammatic Monte-Carlo \cite{simkovic2022}, Neural Network based variational ansatz \cite{carleo2017}, Projective Entangled Pair States (PEPS) \cite{corboz2010} and more. There are also new types of quantum inspired algorithms that are being developped. For instance, the quantum Fourier transform at the core of Shor's algorithm can be turned into a classical algorithm that performs Fourier transforms exponentially faster than the usual FFT \cite{chen2022}.
It might very well be that some of the promises made for quantum computers will actually be kept perfectly by classical algorithms running on classical hardware. One could even argue that the real goal of physics is not to calculate this or that number but to unveil those hidden structures.

\subsection*{Acknowledgement}
I am grateful to Miles Stoudenmire and Michele Filippone who provided numerous insightful feedbacks on this manuscript.
Warm thanks to Paola Waintal Saldi and Christoph Groth for proof reading this article.

\bibliography{QuantumHouseOfCards}

\end{document}